# Activation and radiation damage in the environment of hadron accelerators

*Daniela Kiselev*
Paul Scherrer Institute, Villigen, Switzerland

**Abstract**
A component which suffers radiation damage usually also becomes radioactive, since the source of activation and radiation damage is the interaction of the material with particles from an accelerator or with reaction products. However, the underlying mechanisms of the two phenomena are different. These mechanisms are described here. Activation and radiation damage can have far-reaching consequences. Components such as targets, collimators, and beam dumps are the first candidates for failure as a result of radiation damage. This means that they have to be replaced or repaired. This takes time, during which personnel accumulate dose. If the dose to personnel at work would exceed permitted limits, remote handling becomes necessary. The remaining material has to be disposed of as radioactive waste, for which an elaborate procedure acceptable to the authorities is required. One of the requirements of the authorities is a complete nuclide inventory. The methods used for calculation of such inventories are presented, and the results are compared with measured data. In the second part of the paper, the effect of radiation damage on material properties is described. The mechanism of damage to a material due to irradiation is described. The amount of radiation damage is quantified in terms of displacements per atom. Its calculation and deficiencies in explaining and predicting the changes in mechanical and thermal material properties are discussed, and examples are given.

## 1     Activation on a microscopic level

Nuclear reactions of particles with the nuclei of the chemical elements of a component may cause an initially non-radioactive component to become radioactive. In contrast to chemical reactions, where the elemental composition is of interest, the isotopic composition of the component has to be known. An isotope is characterized by its number of protons, $Z$, and its number of neutrons, $N$. The total number of protons and neutrons is called the mass number, $A$. Before exposure to irradiation, the distribution of the isotopes of each element is given by the natural abundance. (Cosmic radiation also produces radioactive isotopes in materials but at a very low level. This becomes relevant, for example, when a material is used as shielding for measurements sensitive to the natural background. In this case ancient lead, which can be found in ancient Roman ships buried in deep water, can be used. The water provides effective shielding against cosmic irradiation, and the radioactive isotopes induced by cosmic rays in ancient times have mostly decayed.) Bombarding isotopes with particles leads to a change in their proton and neutron numbers, i.e. transmutation. When we look at a nuclear chart of isotopes, we see that most of the stable isotopes are directly surrounded by radioactive ones. For interactions with particles of energies larger than 100 MeV, spallation is the dominant process. Part of the energy of the primary particle is transferred to several nucleons in independent reactions. The nucleons inside the nucleus subsequently collide with each other, distributing the energy almost equally. This is called an intranuclear cascade and leads finally to a highly excited nucleus. This process takes about $10^{-22}$ s. Some of the nucleons may leave the nucleus if their energy is higher than their binding energy. Nucleons at the surface, i.e. in less strongly bound states, have binding energies of a few MeV, whereas nucleons in the deeper-lying shells in a medium-heavy nucleus need about 50 MeV to remove them from the nucleus. Charged particles such as protons or pions (produced mainly in the decay of the Delta resonance) have to overcome the Coulomb barrier (the potential energy between the charged nucleus and the particle) in addition. (The Delta resonance is the first excited state of a nucleon, about 300 MeV above the mass of the nucleon. At this energy, a clear enhancement (bump) appears in the cross-section.) Therefore, in the first stage of the spallation process, i.e. the cascade/pre-equilibrium stage, high-energy secondary particles such as protons, pions, and neutrons are emitted. Their energy usually exceeds 20 MeV and most of them move in the direction of the primary particle, which has transferred a high momentum in the first collision to the secondary particles now leaving. In a thick target or component, it is likely that the secondary particles will interact with other nuclei. This is called an internuclear cascade.

In the second stage of the spallation process, the excited nucleus releases its energy by emitting particles. This is called the evaporation phase and takes about $10^{-18}$ s. Mostly neutrons, with a Maxwell–Boltzmann distribution peaking around 1–2 MeV, are emitted. Charged particles such as protons, pions, and light ions are suppressed owing to the Coulomb potential, particularly for nuclei with a large charge number. If the energy left falls below the threshold for particle emission, the remaining energy is released by photon emission. Owing to the deficit of neutrons, the remaining nucleus is likely to be radioactive. For heavy nuclei, high-energy fission competes with evaporation during the evaporation phase and is the preferred channel for very heavy nuclei such as lead, tungsten, and mercury. At high excitation energy, the nucleus breaks into two halves, i.e. symmetric fission is preferred. Some lighter particles such as neutrons and protons are lost during the fission process.

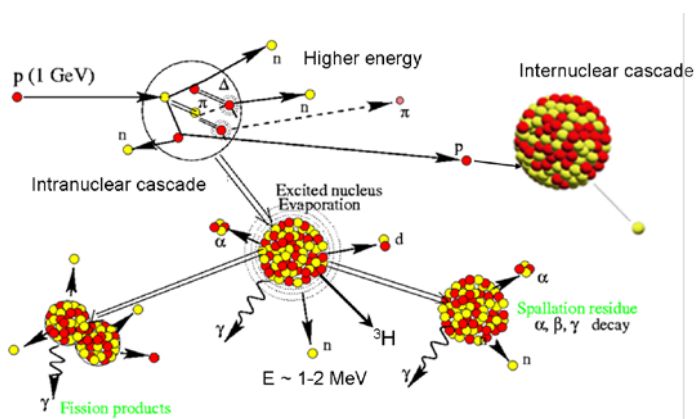

**Fig. 1:** Illustration of the spallation process (modified from an image on www.cea.fr)

The spallation process is illustrated in Fig. 1 for a 1 GeV primary proton. At higher energies, the process is similar. More particles and other species such as exotic mesons can be produced. More and larger fragments can be emitted. In general, the number of neutrons increases with energy. For 1 GeV protons, 20 neutrons per primary particle are produced in lead.

Starting from a target nucleus ($Z$, $A$), all nuclei with mass numbers less than $A$ down to 1 can be produced by spallation. For a heavy nucleus, the mass region around $A/2$ is filled by fission residues. How many isotopes of a given type remain after bombardment of a target nucleus with high-energy particles depends on the energetics of the process and, in more detail, on the structure of the initial and final nuclei. The production rate of an isotope obviously depends on the number of target nuclei $N_A$ and on the number of primary particles $n_i$ incident per second onto the target material. The quantity that describes the transformation of a target nucleus $A$ into a given isotope $Y$ is called the (production) cross-section. Since the energy spectrum $\phi_x$ of particles of type $x$ (e.g. secondary neutrons) is given per incident primary particle and per unit area (1/area), the cross-section has units of area. Owing to the small value of the cross-section, the barn (b), equal to $10^{-24}$ cm$^2$, is often used as a unit. The production rate $P_Y$ for isotope $Y$ is given by

$$P_Y = N_A n_i \int \frac{\mathrm{d}\phi_x(E)}{\mathrm{d}E} \cdot \sigma_{xA \to Y}(E) \cdot \mathrm{d}E \qquad (1)$$

Since the production cross-section $\sigma_{xA \to Y}$ depends on the energy, the spectrum of the particles as a function of energy is needed as well.

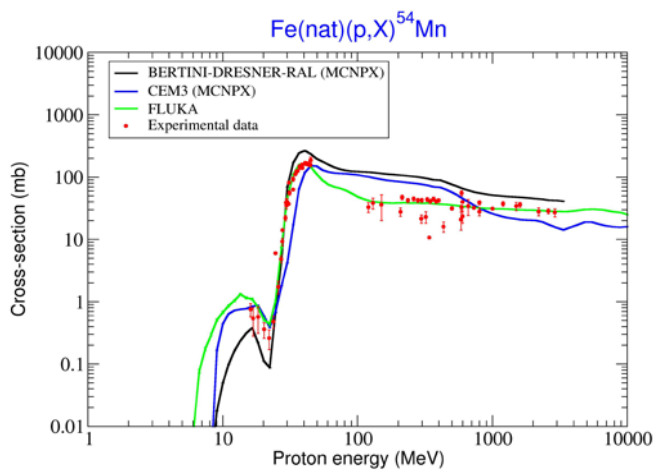

**Fig. 2:** Production cross-section for $^{54}$Mn using proton irradiation of natural Fe. Results obtained from several models [1–6] implemented in MCNPX [7] and experimental data [8] are shown. FLUKA [9] simulation: courtesy of S. Roesler, CERN.

A typical production cross-section for $^{54}$Mn using protons on natural iron is shown in Fig. 2. For protons with energies larger than a few hundred MeV, the cross-section becomes almost constant. This behaviour continues with a slight slope up to the TeV range. A similar behaviour of the cross-section is observed for high-energy neutrons, except at the threshold. Owing to the charge of the proton, a minimum energy is needed to interact with the charged nucleus. This threshold energy depends on the nuclear structure and $Z$. For medium-heavy nuclei, the threshold energy is a few MeV. The bump between 10 and 20 MeV in Fig. 2 is caused by a large number of close-lying resonances, which correspond to nuclear levels that are free to capture a proton.

The curves in Fig. 2 are the predictions of various models implemented in the particle transport codes MCNPX [7] and FLUKA [9]. The models CEM3.02 [5, 6] and BERTINI-DRESNER-RAL [1– 4] were used in the form of versions implemented in MCNPX2.7.c. Compared with the experimental data [8] shown by the points in the figure, all models are in reasonable agreement. This is not always the case, however. In addition, there may be little no data available for elements less common than iron and, particularly, for targets consisting of a single isotope. Models need to predict not only the production cross-sections but also the production of secondaries and their angular distribution. Owing to the interest in low-energy neutrons for spallation sources and the high penetration capability of high-energy neutrons through biological shielding, the prediction of neutron production is the most important requirement in many cases. In addition, the main contribution to the activity is often caused by low-energy neutrons, owing to their large capture cross-sections (see below).

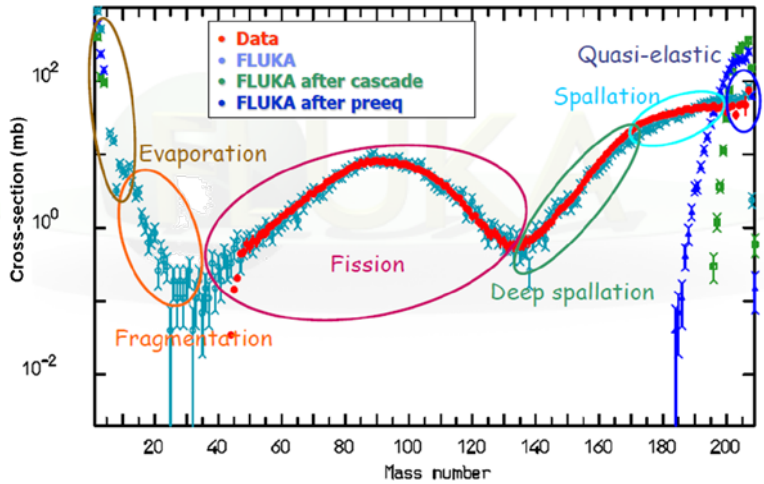

**Fig. 3:** Mass distribution predicted by FLUKA after bombardment of Pb with 1 GeV protons. The data are taken from Ref. [10]. Image courtesy of A. Ferrari, CERN.

Figure 3 shows the mass distribution for 1 GeV protons on $^{208}$Pb, calculated with FLUKA; the inverse reaction was measured at GSI [10]. The FLUKA results are in very good agreement with the data shown as dots. Up to 15 mass units below the target nucleus, the isotopes produced arise from spallation reactions. The production cross-section for these isotopes is quite high. Isotopes which lose more nucleons receive a larger energy transfer from the projectile, and therefore this type of reaction is called 'deep spallation'. Afterwards, the fission products are broadly distributed around half of the mass number of the target nucleus, since symmetric fission is favoured for heavy nuclei at higher energy, with a few nucleons being lost during the break-up. At very low mass numbers, evaporation products up to light ions have a high probability of being produced, since they accompany almost every spallation process. Products with larger mass numbers are produced by fragmentation, where the target nucleus disintegrates into large clusters or one cluster is emitted. This region is particularly difficult to reproduce by models.

Further away from the loss points of the primary particles, i.e. in the biological shielding around these loss points, the charged particles are slowed down by ionization of atoms along their paths. Depending on the type of particle, they may decay, be absorbed, or be changed by reactions. An almost stopped proton, for example, can capture an electron and remain as hydrogen in a component. After a certain distance in the shielding material, which depends on the stopping range of the charged particles and therefore on their energy, only neutral particles are left. Since neutrons have a larger lifetime than other neutral particles, and because they are produced in vast amounts, the reactions occurring in the shielding are driven by neutrons. The neutrons lose energy in many (elastic) collisions, i.e. they are moderated. This process is particularly effective for nuclei with small mass numbers, for kinematical reasons. For example, a ball thrown against a wall bounces back with the same velocity, i.e. it does not lose any energy but just changes momentum. The wall represents a collision partner of large mass number, and the ball represents a neutron.

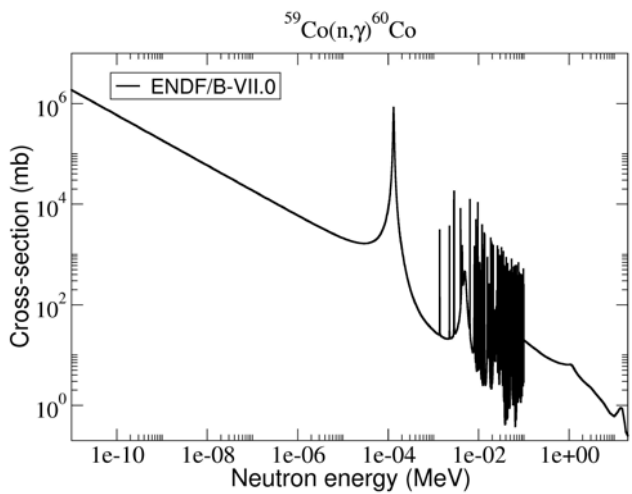

**Fig. 4:** Cross-section for neutron capture by Co for energies up to 20 MeV, from the ENDF/B-VII library [11]

There are still higher-energy neutrons left, but these are in equilibrium with the low-energy neutrons. This means that low-energy neutrons which are lost by absorption are replaced by moderated, formerly higher-energy neutrons. In this case, all energy groups of neutrons are in equilibrium with each other. As a consequence, the shape of the neutron spectrum does not change any more (or at least not significantly); only the amplitude changes, owing to the continuous loss of neutrons. The main reaction process for this is neutron capture, where a neutron is absorbed by a nucleus. As an example, the neutron capture cross-section of $^{59}$Co is shown in Fig. 4. For neutron energies from a few keV onwards, large spikes in the cross-section can be seen. The probability of capturing a neutron with an energy just corresponding to a level of the nucleus that is free to capture a neutron is especially high. This situation is called a resonance. The energy width of the levels is around 1 eV. At higher energies, the density of states increases and the resonances overlap. The neutron is usually captured into an excited state. The surplus energy, including the binding energy, is often released as a photon. For energies less than 1 eV, the cross-section increases linearly on a logarithmic plot, with the inverse of the square root of the energy or the inverse of the velocity. As the velocity decreases, the time for interaction between the neutron and the nucleus increases, and therefore the probability of capture of the neutron increases.

## 2    Direct calculation of the activation

Spallation is a complicated process with many open reaction channels. The particles have to be tracked, particularly in thick targets where secondary particles produce particles again and so on. This is a kind of chain reaction, for which Monte Carlo simulations are useful. The code for the simulation has to provide cross-sections for all relevant reaction channels and particles involved. Usually, the cross-sections for low-energy neutrons up to 20 MeV are taken from libraries, where experimental cross-sections have been cross-checked against each other and combined with models where necessary. A well-known and often used library is one from the USA and Canada compiled by the Cross Section Evaluation Working Group (CSEWG). The newest version available is ENDF-B-VII [11] (ENDF is an abbreviation for 'Evaluated Nuclear Data Files'). The result of the European effort in this field is the Joint Evaluated Fission and Fusion (JEFF) file [12], compiled by the NEA Data

Bank member countries. Only recently have libraries for neutrons, protons, and deuterons up to 200 MeV, such as EAF-2010 [13], become available. Usually, models are used in simulation codes for neutrons above 20 MeV and for all other particles. At present, most of these models are based on the mechanism of the internuclear cascade (INC) described above. Their origin can often be traced back to the 1960s. The INC models are coupled to models for fission and evaporation on a microscopic basis.

The user of the Monte Carlo code has to provide the geometry of the system and the characteristics of the particle beam, which define the primary particles. If the component of interest is far away from the first point of interaction and the space between is filled with material with a complex geometry, the geometric model that must be provided is elaborate and the calculation is CPU-consuming. Often tricks (biasing) have to be applied to get sufficient statistics at the point of interest. In some restricted cases, other methods are more efficient than a Monte Carlo simulation. One method will be presented below.

Besides the geometry, the material composition of all components of interest has to be known. The most important constituents are elements with large production cross-sections, such as Co for neutrons. Only a few thousand parts per million of Co is contained in steel, and a few parts per million in aluminium (depending on the grade). Owing to its large neutron capture cross-section, $^{60}$Co becomes dose-relevant after a cooling time of about a month. In aluminium, the dose is determined by $^{22}$Na initially, and by $^{60}$Co after a few years of cooling. Therefore it is important to know precisely the Co content. In practice, however, the Co content is not precisely known. The definition of the material is one large source of uncertainty in the prediction of the dose. Some examples will be given below.

A pure Monte Carlo simulation follows each particle through the material, including all interactions between the material and the particle. At the end of the simulation, the number of reaction products is counted for each isotope per primary particle, which leads to the production rate for a specific isotope (see Eq. (1)). Alternatively, particle fluxes can be provided, which are later folded with the corresponding cross-sections. To obtain the activity of a component, the time periods during which the component was irradiated have to be known. Cooling times, i.e. periods with no irradiation, are also important, because of the decay of the radioactive nuclei. First, the time evolution of an ensemble of $^{60}$Co nuclei will be considered for simplicity. At the beginning ($t_0 = 0$), there are $N_0$ nuclei in the ensemble; after a time $t$, the number of nuclei that have not yet decayed follows the law of radioactive decay:

$$N(t) = N_0 \exp(-\lambda t) , \tag{2}$$

where $\lambda$ is the decay constant and is related to the half-life $T_{1/2}$ via

$$\lambda = \frac{\ln 2}{T_{1/2}} . \tag{3}$$

The activity is defined as the number of nuclei that decay per second, i.e.

$$A(t) = -\frac{dN}{dt} = \lambda N(t) . \tag{4}$$

The expression on the right-hand side is derived from Eq. (2). During the irradiation time, the number of nuclei of an isotope can change not only by decay but also due to the production rate $P$ of that isotope. The rate equation is then

$$\frac{dN(t)}{dt} = P - \lambda N(t). \tag{5}$$

The expression for the activity follows from the solution of the differential equation in Eq. (5):

$$A(t) = P(1 - \exp(-\lambda t)). \tag{6}$$

For very long irradiation times, the activity is equal to the production rate $P$ and becomes independent of the irradiation time. It is important to realize that at a given particle flux, the activity cannot exceed a certain value, which is called the saturation activity $A_{sat}$. An example of the production and decay of $^{60}$Co is shown in Fig. 5. $^{60}$Co has a half-life of 5.3 years; 90% of the saturation activity is reached after an irradiation time of about three times the half-life, and half of the saturation activity is produced after an irradiation time of the order of the half-life. Since $^{60}$Co decays into stable $^{60}$Ni, the decay chain ends at that point.

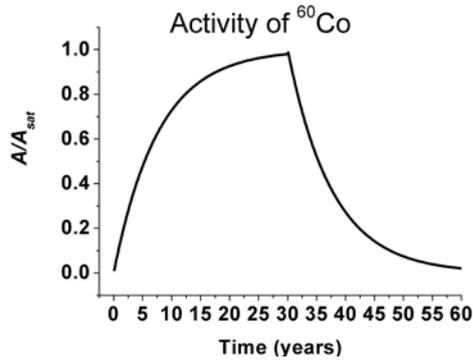

**Fig. 5:** Build-up and decay of $^{60}$Co.

In the general case, many isotopes are produced, each with different production rates. The number of nuclei of an isotope of type $m$ per unit time is given by the Bateman equation,

$$\frac{dN_m(t)}{dt} = -N_m(t)\left(\lambda_m + \varphi_x \sigma_{m+x}^{abs}\right) + \sum_{k \neq m} N_k(t) \varphi_x \sigma_{k \to m} \,. \tag{7}$$

Here the first term describes the decrease in the number of nuclei of an isotope due to radioactive decay and transmutation to other isotopes. The last term contains the production of isotope $m$. The sum runs over all production channels.

There are Monte Carlo codes such as MCNPX and PHITS [14] which have to be coupled to external build-up and decay codes to obtain the activation after a specified irradiation history. "External" means that these codes are not provided by the Monte Carlo distributor. PHITS is often used in connection with DCHAIN [15], and for MCNPX a script [16] exists to transfer the information about the neutron flux (for neutron energies less than 20 MeV) and the residual production rates to Cinder'90 [17] or FISPACT [18]. These programs use a built-in or external library for neutron energies of less than 20 MeV (sometimes reaching to higher energies). They also provide all decay properties of the isotopes. MARS [19] comes with its own build-up and decay code, which is based on DCHAIN. The calculation of the activation for user-specified irradiation conditions is done during runtime. FLUKA provides its own build-up and decay codes. It offers two possibilities, to perform the calculation of the activation during runtime or to do the operation of coupling to the build-up and decay code separately afterwards. The later option needs more effort, but has the advantage that a request for results at more cooling times or different irradiation times after the simulation has been run can be quickly fulfilled.

Owing to the dependence of the production and decay rates on the half-life, the question of which radioisotope dominates the activity in a specified material has a time-dependent answer. It is clear that short-lived isotopes decay rapidly and their presence is negligible after long cooling times. On the other hand, the final nuclide inventory depends not only on the total number of bombarding particles but also on the particle rate (or flux, or current). For example, consider two isotopes with very different half-lives but the same production cross-section. This condition leads to the same saturation activity for both isotopes. The difference is that during a short intense irradiation time, half of the saturation activity is reached for the short-lived isotope, whereas the activity of the long-lived isotope is still negligible. ('Short' means that the irradiation period is of the order of the half-life of the shorter-lived isotope.) When the same number of particles is kept but at a lower flux, the consequence is a lower saturation activity for both isotopes. Compared with the case of the short irradiation time, the activity of the long-lived isotope will be larger. For very long irradiation times, the same activity will be reached for both types of isotopes. This does not mean, however, that the same numbers of nuclei of the isotopes are produced. Since

$$A = \lambda N, \tag{8}$$

the number of nuclei of the short-lived isotope will be greater by a factor of $T_{1/2}$ (long)/$T_{1/2}$ (short).

## 3    Example: activation of a directly irradiated component

'Directly irradiated' means that the component is hit by the primary proton beam. At such locations, a major fraction of the beam is often lost. Such dedicated beam loss points include targets, collimators, and beam dumps. The particular activated component considered in this section is the first version of the target for the neutron spallation source SINQ at the Paul Scherrer Institute (PSI). The 57 cm long target contains about 456 Zircaloy tubes, each of them 10 cm long. The beam hits the tubes, known as 'cannelloni', laterally, i.e. the tubes are installed perpendicular to the beam. Nowadays, these tubes are filled with lead, but in the trial version of the target, solid Zircaloy sticks were installed instead of tubes. From the beginning of the development of the target, the containment around the target served as a 'safety hull'. This containment was made from AlMg3, an aluminium alloy containing 3 weight per cent of magnesium. The target was mounted vertically and the beam hit it from below. Every two years, the target, together with the safety hull and the shielding behind the target, was changed. Samples from the front window of the safety hull, the shielding, and a Zircaloy screw in the target were taken in a hot cell using remote-controlled manipulators. The gamma spectra of the samples were measured with a high-purity germanium detector, and the isotopes present were identified by their characteristic gamma energies. Owing to their high self-absorption, the beta emitters needed a chemical treatment, after which they were dissolved in a scintillating fluid and measured. Isotopes with a very long half-life would need a very long measurement time to achieve the required statistics; these isotopes were instead counted one by one using the technique of Accelerator Mass Spectrometry (AMS). Chemical preparation was needed for this purpose also. The samples were not examined for alpha emitters. The measurement of alpha emitters would need a windowless solid state detector, and the preparation of the samples would be difficult and elaborate.

The geometrical model prepared for the simulation that was performed using MCNPX is shown in Fig. 6, which also gives a good overview of the SINQ facility. Around the target, water (shown in dark grey) is used to decelerate the neutrons to thermal energies. This is similar to the set-up of a reactor. Instead of normal water, the tank is filled with deuterated water, $D_2O$. The reason for this is that hydrogen has a large capture cross-section for thermal neutrons and would therefore reduce the neutron flux significantly. A smaller tank, shown in light grey in Fig. 6, contains $D_2$ at 25 K. Scattering of neutrons by cold deuterons leads to energy transfer from the faster neutrons to the deuterons, resulting in cold neutrons. The intermediate grey areas are made of iron and serve as biological shielding. The concrete shielding outside the iron shielding was not included in the model. The samples were taken from SINQ target 3, which was irradiated for almost two years from 1998 to 1999 with protons equivalent to 6.77 A h. In the MCNPX simulation, the ENDF-B-VI cross-section library was used for neutrons of energy less than 20 MeV. The resulting neutron flux was folded with the cross-sections provided by the Cinder'90 library, version 7.4. The same program also calculated the time evolution of the radioisotopes according to the irradiation history.

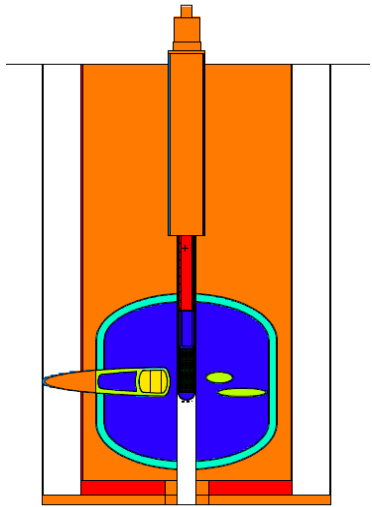

**Fig. 6:** Geometrical model of the SINQ facility at the PSI used for calculations with MCNPX

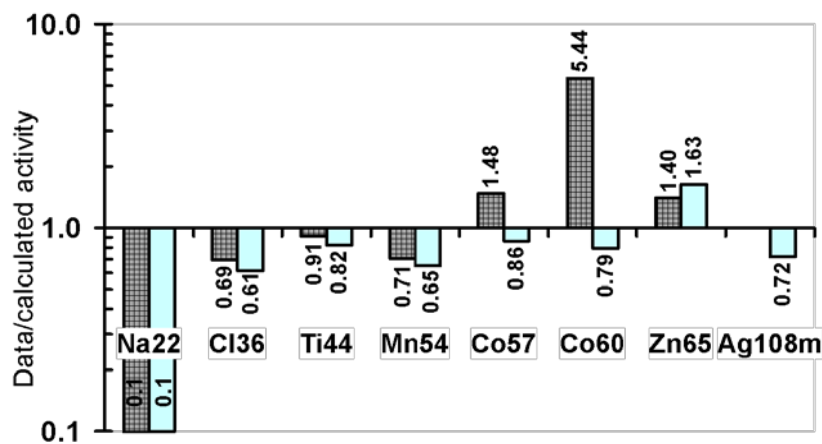

**Fig. 7:** Comparison of the measured and calculated activities of several isotopes produced in the safety hull of SINQ target 3.

The ratio of the measured to the calculated activity is shown in Fig. 7 for several isotopes in the sample from the safety hull. The two shades of grey represent two different material compositions that were used for the calculation. The main differences in these compositions were that the first composition (shown in the darker shade) had a lower Co content and no Ag. As usual, the exact material composition was not known, since an analysis of the material of this particular component was not performed. The concentration of Ag assumed was only 40 parts per million, but this produced a measurable activity of 72 Bq/g $^{108m}$Ag, which is a factor of 10 above the free-exemption limit. This isotope is produced by neutron capture by elementary silver. Therefore no $^{108m}$Ag was obtained with the first material composition, where silver was absent. This example reveals that the material composition is an important input and source of uncertainty when one is calculating nuclide inventories. When the calculated activity is compared with the $^{60}$Co data, the higher Co content in the second material composition fits the measured activity better. Except for $^{22}$Na, good agreement is achieved between the measured and calculated activities. Using a later version of the Cinder'90 library, the measured value of Na-22 could be reproduced almost exactly. The new value was similar to the activity obtained with SP-FISPACT, which leads to a ratio of 0.72.

## 4  Radioactive waste from accelerator facilities

In every country, there are regulations that determine when a component that has been used in a radiologically controlled area can be freely released and when it has to be disposed of as radioactive waste in a final repository. In Switzerland, a (solid) component can be freely released when all of the following three conditions are fulfilled:

1. The dose rate at a distance of 10 cm is less than 0.1 µSv/h. The dose rate is equal to the absorbed energy in 1 kg of human tissue times a biological factor, which takes the damage done to human tissue into account.
2. The sum rule $\sum_i (A_i/LE_i) < 1$ is obeyed, where the $LE_i$ are the exemption limits given in the radioprotection regulations. The sum runs over all radioisotopes.
3. The surface of the component is free of radioactive particles (no contamination) according to the limits defined in the regulations.

Radioactive waste has to be kept for 30 years if there is a chance that the material might be freely released at the end of this storage period or at some earlier time. If this is not the case, components have to be disposed of as radioactive waste. For this purpose, it is required that the documentation accompanying a container of waste contains, among other things, a nuclide inventory of the components in the container. The nuclide inventory should be as complete as possible and should not only provide the main isotopes. However, in practice it is almost impossible to determine all isotopes experimentally. As already mentioned, isotopes which cannot be identified by their gamma spectra are difficult to measure. To maintain confidence in the codes used for calculating nuclide inventories and maintain the acceptability of those codes, comparisons of the calculated results with experimental data are periodically performed. The code used most often for this purpose is called PWWMBS and was developed at the PSI [13]; it will be described in more detail below.

In the last few years, the authorities in Switzerland have also required an estimate of the total amount of future waste, including that arising from decommissioning, when operating approval is being sought for an installation or facility. These calculations usually require a huge effort, since particle transport Monte Carlo calculations are necessary for large regions, sometimes up to the outer biological shielding. This requires biasing techniques, since otherwise only very few particles reach the outer shielding, not enough to make statistically conclusive statements about the nuclide inventory. When the flux consists mainly of low-energy neutrons, it is possible to obtain the activities by folding the flux with known cross-sections using an external program.

The radioactive waste at the PSI is produced mainly at the proton accelerator. The amount of waste from the electron accelerator at the Swiss Light Source (SLS) is negligible. The reason for this is that electrons produce mainly bremsstrahlung and undergo almost no nuclear reactions, although a small probability exists that hard gamma photons will produce neutrons via photonuclear reactions. Compared with a hadron facility, however, the residual activity remains small. In contrast to nuclear power plants, where high-level waste is regularly produced when the spent fuel elements are exchanged, most of the operating waste at the PSI is low-level waste. The majority of components have dose rates below 100 µSv/h. In a few hundred years, the activity of these components will decay to a level where they can be freely released. In France, there are repositories on the surface already in use for waste which needs a cooling time of less than 300 years to be declared as conventional waste. In Switzerland, however, such a storage facility is not foreseen. Here, waste is only separated into constituents that do and do not generate a considerable dose. Fuel elements belong to the category HAA (High Activity Waste), and waste from accelerators to the category SMA (which refers to waste with low and intermediate levels of activity). Alpha-toxic waste, i.e. waste containing an amount of alpha-emitters larger than 20 000 Bq/g, is distributed between repositories for both types of waste according to its dose.

The regular operating waste at the PSI is 90% normal and stainless steel. Typical components are vacuum chambers, beam diagnostic elements, magnets, and shielding, and to a lesser amount targets, collimators, and beam dumps. The large amount of biological shielding made out of steel and concrete will be disposed of at the time of decommissioning of the facility. Owing to the higher energies of neutrons produced at accelerators compared with nuclear power plants, the inner shielding consists of a material denser than concrete; normal steel is usually chosen.

At the PSI, the solid radioactive waste is placed in concrete containers. Finally, the waste components in the container are fixed in place with concrete and the container is closed with a cover. The resulting package is suitable for a final repository. Because the combination of aluminium and concrete is a source of hydrogen, the surface area of aluminium components has to be reduced by melting them down; the molten metal is the cast into coquilles. On average, a container with an opening of size 1.26 m × 1.26 m and a height of 1.88 m can be filled with 4.5 t of waste; the last 30 containers achieved a better filling factor and held an average of 6 t of waste. In the case of aluminium, 36 coquilles can be fitted into a container, which gives a total of 1.8 t. The activity per container is about $10^{10}$–$10^{12}$ Bq. If there is no major reconstruction work at or in the close environment of the proton accelerator, only one container per year is filled.

## 5      Determination of the nuclide inventory for radioactive waste

As already mentioned, a nuclide inventory is required by the authorities before radioactive waste can be accepted for disposal in a repository. To perform a particle transport Monte Carlo simulation for every component in a container would require a huge effort, particularly if a complex geometry needed to be modelled. This is often the case when a component is far away from the loss point(s), i.e. the points where the primary beam interacts with a material to produce a shower of a variety of particles. If the particles have no charge, they can travel large distances and thus activate components far from the source (i.e. the loss point). (Although the muon is charged, its mean free path is large. In contrast to the electron, the main mechanism of energy loss, due to bremsstrahlung, is suppressed owing to its 206 times larger mass; the bremsstrahlung cross-section is proportional to $1/m^2$. As the muon is a lepton, its cross-section for hadronic reactions is small.) Because these loss points are surrounded by numerous components (some of which are listed above) and large blocks of shielding material, most of the radioactive waste in a container has not been directly irradiated by the primary beam. Among the neutral particles, only neutrons and $\pi^0$ particles can be produced from the 590 MeV proton beam available at the PSI. The $\pi^0$ has a lifetime of $8.4 \times 10^{-17}$ s, much too short to cover a considerable distance. Therefore the only significant source of activation far from the loss points is the neutron flux. It should be mentioned that photons also reach far beyond the loss points; these are produced via bremsstrahlung from charged particles. However, they do not play a role, since their cross-section for nuclear reactions is much smaller than that for neutrons.

The basic idea is to use Eq. (1) to calculate the production rate for each isotope by folding the neutron flux spectrum with the corresponding cross-section. This is a fast and simple method, provided that the neutron flux spectra are known. The key to avoiding calculating the neutron spectra at every location by a complex Monte Carlo simulation is the observation that the shapes of the neutron spectra do not vary much within a large region, i.e. the energy dependence is almost constant even though the amplitude varies. The reason is that high-energy neutrons lose energy while, at the same time, low-energy neutrons are captured or stopped. In the ideal case, the high- and low-energy parts are in energy balance.

This method is applied in the code PWWMBS [20], developed at the PSI. The name is an abbreviation of 'PSI West Waste Management Bookkeeping System'. The code contains its own cross-section library for neutrons from 2 MeV upwards, extracted from older models based on INC. This library is called PSIMECX [21]. For neutrons with energies less than 20 MeV, evaluated cross-sections are used. The neutron spectra are calculated under simplified assumptions about the geometrical layout, omitting many of the details. A crucial input is the material composition. As was shown above, small amounts of impurities can produce significant amounts of radioisotopes if the production cross-section is large. The material compositions used are averages over several material analyses performed on different types of materials used at the PSI. Several different methods can be applied to determine the material composition. Gaseous products are more difficult than other elements to determine. Furthermore, detection at the level of a few parts per million is more difficult to achieve for heavy elements.

To determine the nuclide inventory of a component, its irradiation period, location, and weight have to be known. PWWMBS contains a data bank which stores the charge delivered by the proton accelerator to the target stations integrated over one year, as well as the main shutdown periods. The corresponding neutron flux spectrum is chosen depending on the location. Since there is no normalization of the spectrum to the actual neutron flux, the nuclide inventory is calculated in an initial step with an arbitrarily assumed flux. The surface dose rate is then calculated from the resulting inventory and compared with the measured value, and the calculated dose and the nuclide inventory are scaled to the measured dose rate.

PWWMBS is applicable to more than 90% of the waste generated at the PSI. In the case of targets, collimators, and beam dumps, MCNPX or FLUKA is used to determine the nuclide inventory.

At CERN, a code based on the same principles has been developed recently, called Jeremy. Owing to the higher energies of the protons used at CERN, spectra for pions, protons, and photons are included as well as neutrons [22].

## 6    Example: activation of an indirectly irradiated component

In 2004, several samples were taken from the μE4 beam line. This beam line is one of the six beam lines around Target E, which is a graphite wheel rotating at 1 Hz. As a result of reaction of the 590 MeV protons with the graphite, pions are produced, in addition to neutrons. The pions decay into highly polarized muons. Both types of particles are used for experiments in the fields of basic physics and materials science. The μE4 beam line was reconstructed in 2004 with the aim of improving the efficiency with which muons were collected and transported to the experimental area. Many components were removed on this occasion and became radioactive waste. Samples were taken from the bending magnet ASK61, the shutter behind ASK61, and the shielding around the shutter. The location of ASK61 relative to Target E and the main beam line is shown in Fig. 8. Since the components in the μE4 beam line are not directly irradiated, it is justifiable to calculate their nuclide inventory with PWWMBS. We shall present a comparison between the measured and calculated values of the activity of several isotopes for a sample taken from the beam tube in front of ASK61.

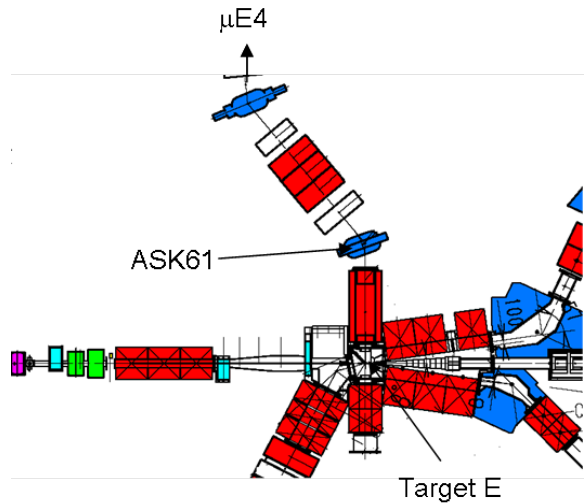

**Fig. 8:** Area around Target E with the μE4 beam line at the top

The beam tube was made from stainless steel and had been in place since 1991, when the Target E area was rebuilt. The sample was obtained by drilling a hole into the beam tube and therefore consisted of swarf. The operation was performed with a manipulator in a hot cell at the PSI. As usual, the activities of several isotopes were measured via gamma and beta spectroscopy. For long-lived isotopes such as $^{36}$Cl and $^{26}$Al, the activities were determined via AMS. Since the gamma measurement was performed as soon as possible after the last irradiation period, isotopes with half-lives of the order of a few months could be detected.

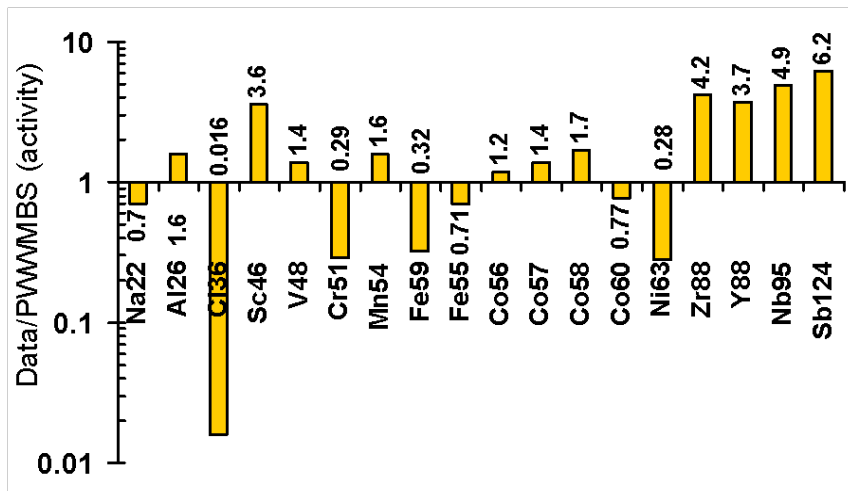

**Fig. 9:** Ratio of the measured to the calculated activity for several isotopes in a sample from the beam tube at the entry to ASK61

The experimentally determined activities of several isotopes are compared with the values calculated with PWWMBS in Fig. 9. The neutron flux spectrum for the region perpendicular to Target E was chosen. The dose rate at the location where the sample was taken was measured to be 70 mSv/h at the surface, two months after end of beam. (In fact, it was not possible to measure the surface dose rate, but only the dose rate at some distance from the surface. This distance was 3 cm for the Geiger–Müller counter used for the measurements. Since the components were large compared with the sensitive area of the counter, the dose rate did not vary significantly with distance.) The results in Fig. 9 show fair agreement between the measured and calculated results, within a factor of 10; this factor is anticipated owing to the uncertainty of the method. An exception is the overestimation of $^{36}$Cl in the calculation by almost a factor of 100. A likely reason for this is that the chloride content in the material composition used for the stainless steel was too large. The standard method of material analysis (inductively coupled plasma optical emission spectrometry) is not sensitive to gaseous elements, and therefore the Cl content was conservatively estimated. This example shows again the sensitivity to the material composition. The latter is an important source of uncertainty.

## 7       Radiation damage in materials

In this second part of the paper, another implication of particle irradiation of materials will be examined. This is the change in material properties due to damage to the lattice structure, which sometimes leads to the failure of components. This effect is called radiation damage. It is a threat particularly to components at loss points in high-power accelerators. These components include targets, beam dumps, and highly exposed collimators. There is renewed interest in the topic of radiation damage owing to new projects and initiatives which require high-power accelerators, and therefore materials which will withstand the high power for sufficiently long. One such project is the European Spallation Source (ESS), which will be built in Lund, Sweden [23]. A rotating wheel made of tungsten with a tantalum cladding is proposed for the target, which will be irradiated with 5 MW of 2.5 GeV protons. Some key values which are important when the behaviour of components under radiation is considered will be given below. The Facility for Rare Ion Beams (FRIB) will be built at the National Superconducting Cyclotron Laboratory (NSCL) at Michigan State University. This will deliver heavy ions with an extremely high power density of 20–60 MW/cm$^3$ [24]. The Daedalus project is an initiative at MIT with the aim of studying CP violation [25]. For this purpose, a neutrino beam is produced by three cyclotrons, each delivering a proton beam with an energy of about 800 MeV. The beam power on the target will be 1, 2, and 5 MW for the first, second, and third cyclotron, respectively. An upgrade to higher beam power is already foreseen. At the PSI, a 1.3 MW proton beam is routinely available, which constitutes the most intense proton source in the world at present. An upgrade to a higher beam power is planned (up to 1.8 MW) in the future.

For these projects, it is essential to know how long the heavily irradiated components will survive. In addition, improvement of the lifetime of components needs knowledge about the underlying mechanism of radiation damage and its relation to the changes in material properties. One problem is that the components cannot be tested under the same conditions as the ones they will be exposed to when the facility is in operation. Therefore the correlations between data obtained under different conditions need to be understood.

The macroscopic effects on structural materials caused by radiation damage are the following:

– hardening, which leads to a loss of ductility;

– embrittlement, which leads to fast crack propagation;

– growth and swelling, which lead to dimensional changes of components, and can also induce additional mechanical stress;

– increased corrosion rates, in particular in contact with fluids;

– irradiation creep, which leads to deformation of components;

– phase transformations in the material or segregation of alloying elements, which leads to changes in several mechanical and physical properties.

A component heavily irradiated by charged particles often needs cooling, since the charged particles deposit additional energy. Predictions of the temperature distribution in a component rely on a knowledge of the thermal conductivity of the material. Unfortunately, the thermal conductivity is subject to change. Its precise behaviour is not known for all conditions and materials. Owing to the damage to the lattice structure, one can expect that the thermal conductivity of a high-conductance material will decrease. This has been confirmed for most materials. The consequence is that the component might reach higher temperatures than foreseen, which could lead to the failure of the component. Since the thermal conductivity is difficult to measure, the electrical resistance can be determined with a nanovoltmeter. The two quantities are related to each other by the Franz–Wiedemann law, which contains a phenomenological constant, called the Lorenz factor; this varies for different materials and depends slightly on the temperature.

With pulsed sources of charged particles, components suffer from thermal cycles, which might lead to fatigue. Cracks may occur, which could lead to failure of the component. This phenomenon is also influenced by radiation. Thermal shock is best absorbed in materials with a low thermal expansion. Therefore the thermal expansion serves as a key parameter when one is examining materials after irradiation. Moreover, a drastic change in the thermal expansion with temperature is a clear sign of a phase transformation.

In the following, some examples of observations of radiation damage will be given. In preparation for the above-mentioned FRIB, several objects were studied at the NSCL with respect to radiation damage due to heavy ions. For this purpose, a 580 mg/cm$^2$ tungsten target, which corresponds to a thickness of 0.03 cm, was irradiated with $^{76}$Ge$^{30+}$ ions (which means that only 30 electrons were stripped off the Ge atoms) at an energy of 130 MeV/nucleon. The total energy of the ions was 130 MeV × 76, which leads to 9880 MeV. After irradiation of the tungsten foil with $5.77 \times 10^{16}$ Ge ions on a beam spot with a diameter of 0.6–0.8 mm, a crack was observed. In Fig. 10, it can be seen that this crack is centred on the beam spot, where it has a small bow. Further investigations [24] revealed that the crack was caused by swelling and embrittlement, which induced stress in the foil. The stress might have been increased by thermal stress due to a decrease in the thermal conductivity.

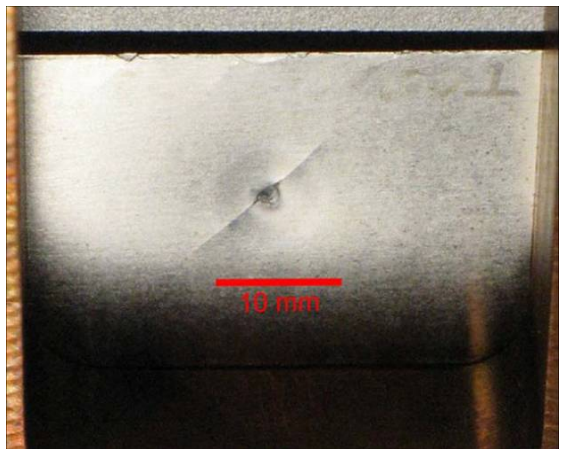

**Fig. 10:** Tungsten foil irradiated with $^{76}$Ge$^{30+}$ ions at the NSCL. Image taken from [24]

At the Los Alamos National Laboratory, tungsten was considered as a material for spallation targets. To investigate its suitability, hardness and compression tests were performed at room temperature and at 475°C on irradiated and unirradiated specimens. Tungsten rods were irradiated for up to 6 months with 800 MeV protons at a current of 1 mA, which corresponds to a dose of 23

Displacements Per Atom (DPA). The temperature during irradiation was kept constant for each sample; it varied between 50 and 270°C for different samples. In the compression tests, the samples were compressed to a strain of about 20%. The irradiated samples suffered from a loss of ductility, which showed up in the compression tests as a longitudinal crack, i.e. in the direction of the force. The compressive yield stress and the hardness increased linearly with the dose after a strong increase at small values. Optical micrographs of the tungsten compression specimens were also taken [26].

The pyrolytic graphite target at TRIUMF was cooled on the edges with water. After irradiation with 500 MeV protons at a current of 120 μA, the graphite delaminated, i.e. segmented into slices perpendicular to the beam. It is interesting that the target survived currents below 100 μA but never above that value. For details and a picture of the target after irradiation, see [27].

## 8    Underlying mechanism of radiation damage

To understand the mechanism of radiation damage, we have to take a look at the various interactions of particles with the atoms of a material. When particles penetrate into a material, they lose energy by several different mechanisms. These are

– electronic excitations;

– elastic interactions;

– inelastic reactions.

The first of these types of interaction is due to the Coulomb interaction and is therefore possible only for charged particles. Here, energy is used to shift electrons from the atomic core to an outer shell. This is called excitation and can also lead to the removal of an electron, i.e. ionization of the atom. The excess energy is dissipated as heat. In some cases this heat might also cause damage to a structural material, although this kind of damage has nothing to do with radiation damage. In elastic and inelastic interactions, energy and momentum are transferred from the particle to the nucleus. In the case of an elastic interaction, the nucleus is not changed but remains the same isotope. In all cases, the atom gains a recoil momentum. If the recoil energy exceeds a threshold (see below), the atom can leave its lattice place. Since the energy and momentum are transferred to the nucleus and not to the electrons, the atom moves in a partly ionized state through the lattice. The recoil energy is lost mainly by Coulomb interactions (ionization and excitation) and is again dissipated as heat. If the energy is large enough, the primary atom can knock on another atom, which again leaves its site. As a result, many atoms can be moved from their original lattice sites. This is called a displacement cascade. The third type of interaction, inelastic reactions, was described earlier in this paper in the part dealing with activation. This leads to transmutation of the nucleus, which can become radioactive. The transmuted nucleus, referred to as an impurity in the following, does not fit ideally into the lattice structure and therefore changes the mechanical properties of the material. Furthermore, in high-conductivity materials such as very pure copper, impurities are known to reduce the conductivity, i.e. they also have an influence on the physical properties. Usually, the damage done to the lattice by the recoils is much larger than that due to the impurities. An exception is when large amounts of helium and hydrogen are produced in highly energetic reactions. In particular, helium can have a considerable influence on the change in material properties, which is not yet fully understood. This will be discussed below.

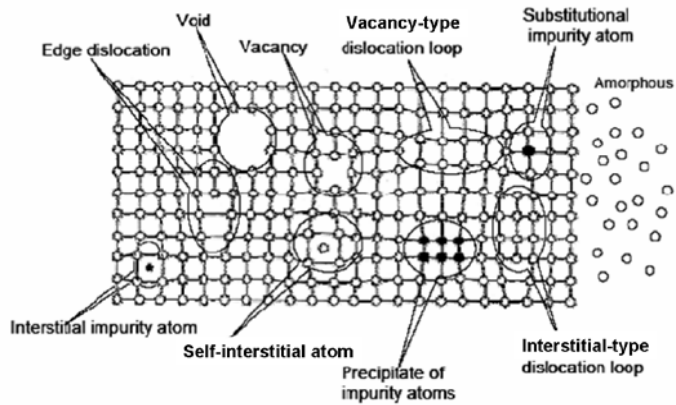

**Fig. 11:** The most important defects in a lattice structure (modified version of an image from Professor H. Föll, University of Kiel).

The most important defects in a lattice are shown in Fig. 11. The open dots belong to the original crystal, and the black dots indicate impurities. The simplest defects are the point defects, also known as zero-dimensional defects. The most prominent representatives of the point defects are self-interstitials and vacancies. Self-interstitials are atoms from the lattice which have left their lattice site for a site not provided in the lattice. The influence of a self-interstitial on its surroundings is a shift of neighbouring atoms away from the self-interstitial to make space for it. A vacancy is just the opposite of a self-interstitial. Here, a lattice atom is missing. These defects also exist in unirradiated materials. If a defect of this type is caused by irradiation, a vacancy and a self-interstitial appear in a pair. This is called a Frenkel pair. Also, an atom on an interstitial site may have been transmuted by an inelastic reaction to an impurity. It is then called an interstitial impurity atom or extrinsic interstitial.

The dislocation loop belongs to the class of one-dimensional defects. Here, part of a lattice plane is missing or has been added. There are two types of dislocation loop: the vacancy-type dislocation loop and the interstitial-type dislocation loop. In the vacancy type, part of a plane of lattice sites is missing. In the interstitial type, part of a plane of additional atoms has been incorporated into the lattice structure. Dislocations move under the influence of external forces, which cause internal stresses in a crystal. In the ideal case, dislocations move out of the lattice. If more than one plane is involved, a cluster is formed. If several planes are partly missing, one has an agglomeration of vacancies. This is called a void. An agglomeration of impurity atoms replacing neighbouring lattice sites on more than one plane is called a precipitate. Owing to their different sizes and properties, the neighbouring atoms are slightly shifted from their original positions. All of these defects make the lattice less flexible against strain, which manifests itself in a loss of ductility and an increase in hardness. In addition, the material becomes brittle. Small cracks can develop, which may grow further, and this can lead to the failure of a component.

Usually, the interaction with a particle with an energy of more than a few MeV does not cause single defects of the kind described above; instead, a large region containing millions of atoms is affected. For example, a nucleus in gold with a recoil energy of only 10 keV destroys the lattice structure in its surroundings within a radius of about 5 nm. This is called a displacement spike and happens within 1 ps. Since a huge number of atoms is involved in the process, a simulation via a Monte Carlo technique needs considerable effort and a large amount of computer power. Such a simulation has to solve the equation of motion for all atoms at the same time, since each atom can interact with and be influenced by all the other atoms. This is a multibody problem, and the computer time needed grows with the square of the recoil energy of the first knock-on atom. The higher the recoil energy, the greater the number of atoms that are involved. Therefore such calculations are limited to recoil energies less than 100 keV for practical reasons. In addition, the simulation has to be repeated for every recoil energy. This kind of calculation is called Molecular Dynamics Simulation (MDS). The advantage is that the results are quite realistic, and the various kinds of defects produced can be studied in the simulation. The MDS method is the only way to evaluate how many defects disappear as a result of recombination with other defects (see below). Unfortunately, the MDS method can follow the process for only a few picoseconds, whereas the healing process can last for months (again, see below).

A faster but less accurate method is the Binary Collision Approximation (BCA), where only collisions between two (hence the name 'binary') atoms are considered. The other atoms are considered as spectators. The particles are followed via trajectories as in a Monte Carlo particle transport program. This calculational method is much faster than the MDS method and also works well at higher energies. However, when such approximations are made, much less information about the process and the state of the lattice is available compared with the MDS method. For example, no statements about the healing of defects can be made.

To estimate and quantify the severity of the damage, a phenomenological approach was developed by Norgett, Robinson, and Torrens, which dates back to the 1970s [28], known as the NRT model after the authors' initials. To quantify the radiation damage, a value is chosen which indicates how often each atom is displaced on average during the irradiation. This quantity is called the DPA, and is obtained by convolution of the energy-dependent particle fluence $\phi(E)$ (in units of cm$^{-2}$) with the displacement cross-section $\sigma_{disp}(E)$:

$$DPA = \int \sigma_{disp}(E) \, \frac{d\phi(E)}{dE} \, dE \tag{9}$$

The displacement cross-section gives the number of displacements per primary (bombarding) particle or secondary particle (neutrons, protons, etc.). It is a function of the energy of the particle responsible for the damage. The displacement cross-section is obtained from the damage cross-section $\sigma_{dam}(E, E_R)$. This damage cross-section is, in addition, a function of the recoil energy of the Primary Knock-on Atom (PKA), i.e. the atom which was first knocked on by the particle. The PKA displaces other atoms if its recoil energy is large enough (generating a displacement cascade). How many atoms are displaced is given by the damage function $\nu(E_R)$, which is the ratio of the energy available to the energy required for displacing atoms. The energy available for the displacement cascade is called the damage energy, $T_{dam}$. This is equal to the recoil energy minus the energy $E_e$ dissipated in ionization and excitation of the atom. For recoil energies larger than 10 keV, most of the energy is lost by ionization. The fraction of the recoil energy left for the damage energy is called the partition function or, sometimes, the damage efficiency. To displace an atom, energy is required to break bonds. The amount of energy required is roughly twice the sublimation energy because, at the surface, only half of the bonding needs to be broken. In Cu, the energy needed ranges from 18 to 43 eV, depending on the crystal orientation [29]. In most calculations, the effective threshold energy $E_D$ for copper is taken equal to 30 eV, but sometimes 40 eV is used. When the recoil energy $E_R$ is larger than $E_D$ but less than $2E_D$, just one atom can be displaced. The PKA may be captured on the lattice site of the second atom. Since for $E_R = 2E_D$ only one atom is effectively displaced, the damage function is given by

$$\nu(E_R) = \frac{\kappa T_{dam}}{2E_D} \tag{10}$$

The factor $\kappa$ is set to 0.8. It compensates for forward scattering, which will not lead to a displaced atom, owing to its low energy transfer. For $E_R > 2E_D$, a cascade of collisions and displacements will take place.

The displacement cross-section is obtained by folding the damage cross section with the damage function:

$$\sigma_{disp}(E) = \int_{E_D}^{E_{max}} \frac{d\sigma_{dam}(E, E_R)}{dE_R} \nu(E_R)\, dE_R \tag{11}$$

The integration runs over all recoil energies from the threshold, i.e. $E_D$, to the maximum possible recoil energy. The damage cross-section is not a reaction cross-section but is related to the recoil energy spectrum $w(E_R)$ of the nuclei. It states how many nuclei can be found with a certain recoil energy. It is obtained from

$$\frac{d\sigma_{dam}(E, E_R)}{dE_R} = \frac{dw(E, E_R)/dE_R}{xN_V}, \tag{12}$$

where $x$ is the thickness of the sample and $N_V$ is the atom density in atoms/cm$^3$. To obtain the recoil spectrum, the cross-sections of all reactions occurring in the material have to be known. Since Monte Carlo particle transport programs contain models for all nuclear reaction cross-sections over a wide energy range, a popular application of these programs is to use them to obtain the recoil spectrum. Here, it is important to use a thin target to avoid significant energy loss of the primary particle in the sample. If the object of interest has larger dimensions, the recoil spectrum has to be calculated for different energies of the primary particle to account for the energy loss of that particle. This requires several Monte Carlo runs. Furthermore, the fluence of the particles in the object of interest has to be obtained, which is a standard option in such codes. The DPA value is calculated from these two quantities via Eq. (9). Nowadays most Monte Carlo particle transport programs, such as FLUKA, PHITS, and MARS, already have a built-in option to obtain the DPA in one run. This is very convenient and avoids a larger effort.

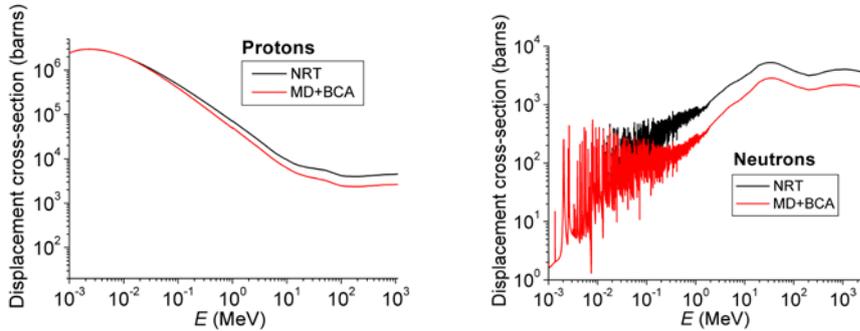

**Fig. 12:** Displacement cross-sections of protons (left) and neutrons (right) in copper obtained by two different approaches (see legend).

As an example, displacement cross-sections in copper are shown in Fig. 12 as a function of energy for protons and neutrons. These cross-sections were extracted by Konobeyev *et al.* [30] using two different approaches: the NRT model, and a combination of the MDS and BCA methods. Owing to the large computer power needed, the MDS method could only be used for energies less than 28 keV. For larger energies, the computation was continued using the BCA. Having these cross-sections, the DPA can be obtained simply via Eq. (9). For thin objects, where the primary energy is roughly constant and the fluence of secondary particles is small, the integration over the energy can be dropped. The displacement cross-sections for protons and neutrons are quite similar in size and shape except at small energies. There, the Coulomb interaction of the proton and the large capture cross-section of the neutron are the main drivers. Above the pion threshold at about 150 MeV, the cross-section is almost constant. This is due to the total inelastic reaction cross-section, which shows the same behaviour for the two particles. The increase in the cross-section at smaller energies is due to the large elastic cross-section, which means that the particle uses its energy most efficiently to displace atoms and not for the release of particles as in the inelastic case.

It should be emphasized that the NRT approach is a simplified method. It completely neglects the details of the process of the displacement cascade. No interactions of the struck atom with the remaining lattice atoms are taken into account. Parameters of the crystal lattice such as the atomic bonding energy and the properties of the solid are completely absent. Instead, all this is condensed into the displacement threshold energy $E_D$. In the NRT model, it is implicitly assumed that the defect concentration is equal to the calculated number of displacements. Moreover, the displacements formed are taken to be stable. Molecular dynamics simulations have shown that the defects are not isolated Frenkel pairs as assumed in the NRT model, but are concentrated in a small region and influence each other. A high density of displaced atoms is produced in the first few tenths of a picosecond. This is called the collisional phase. In this phase, the number of displaced atoms is in fact much larger than that predicted by the NRT model. A few picoseconds later, most of the displaced atoms have recombined with vacancies. This is called 'healing'. The interstitial–vacancy annihilation process is completely omitted in the NRT model. This process is especially important at large PKA energies (>5 keV), where cascades of displaced atoms are produced in the initial state and defects are produced close to each other. At higher PKA energies (>20 keV), subcascades are formed and the number of recombination events decreases. Such a high-energy atom shakes the whole lattice and also deposits thermal energy, localized in the defect region. This makes the defects more mobile and facilitates recombination. The assumption of the NRT model that it is sufficient to count the initially produced Frenkel pairs cannot be justified at energies larger than 0.5 keV, where high-energy cascades start to develop. An example of the effect of healing in copper is shown in Fig. 13. The effective healing is just $1 - \eta$, where $\eta$ is the defect or cascade efficiency. This is given as a function of the recoil energy of the PKA. The defect efficiency $\eta$ is defined as the ratio of the number of Frenkel pairs at the end of phase 1, i.e. at the end of the collision cascade, obtained by an MDS, to the number obtained from the NRT model [31]. The MDS calculation here was performed for a temperature of 4 K. The results confirm that the NRT approach is only justified at small recoil energies. From 5 keV onwards, the recombination of defects dominates. The number of Frenkel pairs is five times lower than that predicted by the NRT model. It is interesting to note that the defect efficiencies in other materials such as W, Fe, and Al show very similar values, even though the final distribution of the defects is different.

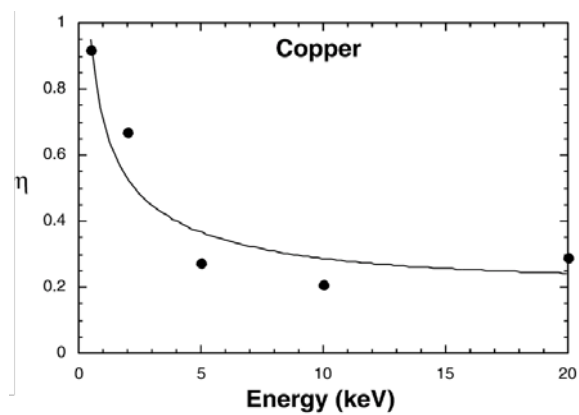

**Fig. 13:** Defect efficiency as a function of the recoil energy in copper [31] at 4 K.

The reduction of the defect efficiency at high PKA energy is important when one is comparing the damage produced by low- and high-energy particles. The materials that suffer radiation damage at today's accelerators are irradiated by high-energy particles, whereas most of the material studies were done in reactors. Figure 13 suggests that one has just to multiply the recoil spectrum of the PKA by the defect efficiency to compare the results and to profit from the large data set that has been taken with reactor neutrons. A comparison of experimental data taken at BNL, where the electrical resistance was measured in irradiated samples of Cu and W, shows that the truth is somewhere in between the predictions obtained by the MDS and NRT methods [31]. The good news is that the NRT–DPA method provides a conservative estimate for predicting the lifetime of a component.

This was also observed in the case of a Cu collimator, called KHE2, in the 590 MeV proton beam line at the PSI. This collimator is located about 5 m behind a graphite target wheel 4 cm thick, which enlarges the beam spot size significantly owing to multiple scattering of protons in the target. The collimator is needed to reduce the beam size, and this also reduces the losses along the beam line. With a 2 mA proton beam on the graphite target, about 150 kW is deposited in KHE2, which has to be cooled with water. KHE2 is 30 cm long, with an outer diameter of 26 cm, and consists of six segments with an elliptical aperture (the half-axes are approximately 4 cm and 8 cm). In the first segment, within a radius of 10 cm, the DPA value obtained with MCNPX was 17.0. Using the NRT cross-sections for neutrons and protons from Konobeyev *et al.*'s work [30] (see Fig. 12), the DPA result was 13.2. At the inner sides of the collimator close to the beam, the DPA value was much higher. The radial distribution of the DPA was calculated with MARS15; the average value found for the DPA was 31.4. The value closest to the beam was about 150 DPA, i.e. five times higher. Although there was a problem that different calculations, all based on the NRT model, led to different results, the DPA at the inner side of the collimator was expected to be at least 80 DPA. In material studies in fast and thermal reactors, a linear volume increase of 0.5% per DPA was measured in several experiments at a temperature of about 400°C, which corresponds to the temperature at the inner side of the collimator at a current of 2 mA. The volume increase, also known as swelling, was measured up to a DPA value of 100, where the volume continued to increase linearly and no saturation was observed (see Ref. [32] and references therein). If one assumes 80 DPA in a volume of 1 cm$^3$, then the opening of the collimator is expected to have shrunk by 1.2 mm on each side. To clarify the situation, the collimator was taken out of the beam line during a long shutdown period of the accelerator. Owing to the high dose rate from the collimator, of up to 300 Sv/h, the handling had to be completely remote-controlled, and additional shielding was provided by an exchange flask (see M. Wohlmuther's contribution to this volume). The aperture was measured with two calibrated laser distance meters. The deviation between the expected and measured values was less than 0.2 mm. The accuracy of the measurement was estimated as 0.5 mm. Images taken with a high-resolution camera revealed no obvious or serious damage to the inner and outer sides of the collimator. Details of the measurement and results can be found in Ref. [33].

The DPA value calculated with the MD/BCA displacement cross-sections shown in Fig. 12 is 7.1. This would lead to 35 DPA at the inner side of the collimator, or a shrinkage of the opening by about 1 mm. This would still be measurable with the laser system used for the distance measurement.

The number of stable defects is reduced at higher temperatures. Two effects contribute to the healing of defects. First, the defect efficiency shown in Fig. 13 was obtained at 4 K, but at higher temperatures it is reduced. According to Ref. [34], the number of defects that survive the collisional phase at room temperature and higher is reduced to about 30% of the value at zero temperature. The reason is that the defects are more mobile at higher temperature. This also helps to reduce the number of defects in the period following the collisional phase; this is the second effect. At room temperature, interstitial clusters migrate to dislocations or recombine with vacancies. This reduces the total number of defects. Vice versa, vacancies also migrate and recombine with interstitials. Large vacancy clusters become mobile at about 400 K. The migration of defects takes time. At room temperature, 25% of the defects surviving the collisional phase disappear after 1 s. After a few months, 50% of the defects have been healed [31]. This effect has been confirmed experimentally with samples which were stored for longer times at room temperature. At higher temperatures, the healing effect would be even larger.

This might be an explanation for why so little damage was observed in KHE2. First, the operating temperature is about 350–400°C at the inner surface. Second, after eight or nine months every year, there is a shutdown period in which the defects have time to recombine.

Many experiments have been done to study the effect of heating samples after irradiation. This heating process is called annealing, and is used for treatment of materials before machining. In these experiments, physical and mechanical properties of samples are measured before and after irradiation, with and without heat treatment. It turns out that part of the damage can be restored, i.e., after the additional heat treatment, the measured values are closer to the original values.

Despite the uncertainty in the prediction of the production of defects in KHE2 and about the possible healing effects during and after the irradiation period, it is questionable whether the swelling rate obtained for irradiation with fast neutrons is the same as that for 590 MeV protons. Irradiation studies at high-energy accelerators are scarce. Some studies on aluminium can be found, because in the 1990s it was a candidate material for the inner wall of ITER. Fortunately, Al has the same lattice structure as Cu (faced-centred cubic, fcc), and therefore the results regarding swelling under irradiation with 600 MeV protons will be summarized here. Details and further references can be found in Ref. [35]. The experiments were carried out at the PSI, which had at that time a station (PIREX) dedicated to irradiation studies. Al foils were irradiated at 120°C with a damage rate of $3.5 \times 10^{-6}$ DPA/s. Samples with doses between 0.2 and 5 DPA were produced. Their microstructure was examined by Transmission Electron Microscopy (TEM), which revealed helium bubbles, as well as voids and other defects. The helium production rate per DPA was about 200 appm (atomic parts per million), which is mentioned here because helium influences the swelling rate (see below). A surprising result [35] was that void formation was observed only at the grain boundaries, and occurred in a specific heterogeneous fashion. No voids were visible in the grain interiors. The original (unirradiated) grain size of the Al samples used was about 200 μm. The observations were completely different from observations on samples irradiated with electrons and neutrons, where there was no lack of voids in the grain interiors. In the TEM images of the proton-irradiated samples, large voids about 50 nm in size could be distinguished from small voids (~20 nm). It is thought that the large voids were nucleated by residual gas atoms, whereas the small voids were produced around helium atoms. Even the large voids in the proton-irradiated samples were much smaller (by a factor of about 3) than the voids produced by fast neutrons. The most important results for solving the 'KHE2 puzzle' came from a comparison of the swelling rates obtained for 600 MeV protons and fast neutrons. The swelling rate as a function of DPA is shown in Fig. 14. For small DPA values of the order of a few tenths, the swelling rates are the same for neutrons and protons. The proton data shown were taken only at the grain boundaries, i.e. in the peak zone (PZ), where the amount of void formation was largest. As already mentioned, no voids were observed in the grain interiors (GI), and therefore the swelling there was zero. At 1 DPA, the difference in the swelling rates in the peak zone for materials irradiated with protons and with neutrons is a factor of 10. Since there were no voids in the grain interiors, the macroscopic swelling rate during 600 MeV proton irradiation was almost negligible compared with that for neutron irradiation. If these findings can be confirmed for Cu also, this would be the key to understanding the good condition of KHE2 after 20 years of heavy proton irradiation.

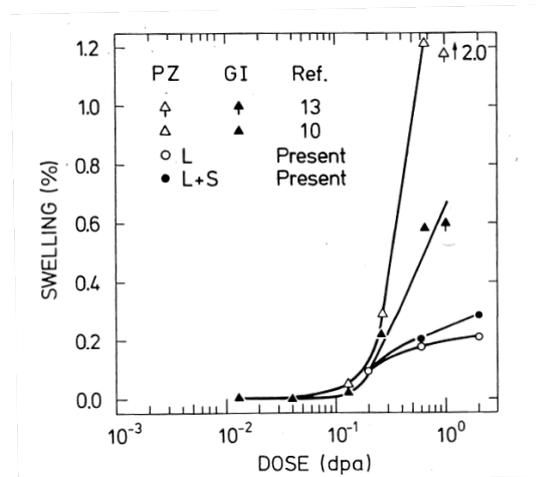

**Fig. 14:** Swelling of Al for irradiation with 600 MeV protons (circles) and fast neutrons (triangles) in the peak zone (PZ) and the grain interiors (GI). The proton data are for the peak zone only. L, large voids; S, small voids. Image reproduced from Ref. [35]. (The column headed 'Ref.' is not relevant to the present paper.)

Another approximation in the NRT model not yet mentioned is obviously the factor $\kappa$, which is taken to be constant in the original NRT model. In MARS, the displacement efficiency $\kappa$ depends on the recoil energy. The values are chosen according to Ref. [36]. For example, at 0.1 keV, $\kappa$ is 1.4. For higher recoil energies, $\kappa$ drops to 0.3 at 100 keV. This modification effectively takes the defect survival into account. The value above 1 for $\kappa$ is due to the fact that the NRT model results used for comparison were calculated with $T_d = 40$ eV instead of 30 eV. In [36], a very weak dependence of the defect efficiency on the temperature and the material was found. These studies were performed with the MDS method.

In their newest versions, the DPA calculations done with MARS and PHITS agree quite well in several cases [37, 38]. According to the authors of those studies, the reason for the agreement and the discrepancy with other codes is the correct treatment of the Coulomb interaction. For highly ionized heavy atoms, the Coulomb interaction is the main driving force up to a few GeV/nucleon. For protons, this is true only for energies up to 20 MeV. Above this energy, inelastic reactions are possible and a lot of secondary particles are produced. Then, mainly secondary particles interact with the atoms and transfer their recoil energy. Since the secondary particles have lower energies than the primary particle, the Coulomb interaction might again be important (depending on the set-up). However, the deviation of the MARS result for the DPA by a factor of two from the result obtained with Konobeyev *et al.*'s displacement cross-sections of cannot be understood in this way. These displacement cross-sections were derived with an explicit treatment of a screened Coulomb interaction. The screening is due to the electron cloud around the nucleus, which effectively reduces the charge of the nucleus for particles approaching the atom. This is only relevant for particles with energies of less than about 2 MeV, since faster particles come close to the nucleus. Moreover, the displacement cross-section for protons incident on Cu given by Konobeyev *et al.* agrees well with the evaluation of Jung [39] in the energy range from 0.5 keV to 20 MeV, which is critical with respect to the Coulomb interaction. Jung deduced cross-sections from measurements of the electrical resistivity at liquid helium temperature under different particle irradiations. Since the change in the electrical resistivity per Frenkel pair defect is known from X-ray scattering, measurements of the electrical resistivity can be related to the number of defects (displacements). More irradiation experiments, for example with neutrons and high-energy protons and a thin-plate target, where no secondaries are produced, are needed to resolve the discrepancy.

Regardless of the details of the calculation and the different results, however, the DPA value predicted by the NRT model is obviously too high and does not reflect the actual number of defects in the material. Unfortunately, the DPA value calculated in the NRT model cannot be measured directly. The reason is that only a small fraction of the displaced atoms lead to permanent lattice defects, as already explained. The state of defects estimated by the NRT model is that after a few picoseconds. The advantage of the NRT–DPA method is that calculations can be done more easily and much faster than MDS calculations. It serves the purpose of comparing and quantifying radiation damage induced in materials irradiated under slightly different conditions. If the irradiation conditions are significantly different, i.e. if the mechanism of defect production changes, the NRT–DPA values are not useful for comparison any more. This is the case, for example, if the lattice type of the material is different or the particles are changed in such a way that the interactions are different.

Heavier ions have a much shorter stopping range in materials than do light ions or protons. This means that the concentration of defects is much higher for the same kinetic energy, and the defects are more localized. At lower energies, it is possible that only the surface of the material might be damaged. Owing to the larger mass of the ions, their energy can also be transferred more efficiently to the target nuclei. At higher energies, the heavy ion rather than its secondary particles is the particle that transfers energy to the target nuclei, i.e. inelastic reactions are negligible. Up to 1 GeV/nucleon, the Coulomb interaction is the most important interaction. Owing to the dominant Coulomb interaction and the high charge, the damage cross-section for a U ion is four orders of magnitude higher than that for a proton at the same energy.

So far, the difficulty of predicting the number of stable defects has been discussed. However, the final number of stable defects may be interesting to scientists who wish to learn more about the underlying mechanisms. In practical cases, the macroscopic effects on the material properties matter. The main changes in materials which can occur under irradiation were listed at the beginning of this paper. The prediction of the behaviour of material properties under irradiation is itself difficult. Much more challenging is the prediction of when this will lead to the failure of a component. The influence of radiation on materials depends on many parameters, such as the rate of irradiation and the type of particle causing the damage. Another category of relevant parameters deals with the structure of the material. These include the grain size of the material and the presence of boundaries, and impurities which have an influence on the lattice. These impurities might be present in the material from the beginning or might be produced during irradiation by inelastic interactions. In addition, the temperature plays an important role. On the one hand, it determines the rate of defect recombination, i.e. healing, after irradiation. On the other hand, the temperature during irradiation also determines which material properties are subject to change. To make the following statements more general, the temperature will be expressed as a fraction of the melting temperature $T_m$ in kelvin. Hardening, i.e. the loss of ductility, occurs after a few tenths of a DPA at temperatures less than $0.4T_m$, and then saturates. From $0.2T_m$ to $0.4T_m$, the material may creep and become deformed. This requires some mobility of the defects, which increases with temperature. Swelling due to voids or vacancies is expected at around $0.5T_m$. Among the impurities produced during the irradiation period, helium is the most important. Since helium is mobile, it migrates into vacancies, which leads to their stabilization. Helium can also accumulate easily in metals, since it is not soluble. It can then form bubbles, which grow with increasing temperature. The result is that the metal becomes brittle and cracks can develop. Only a few atomic parts per million of helium is sufficient to lead to drastic changes in the material properties. Many experiments have performed, some using helium implantation, to study these effects.

Since spallation reactions are often accompanied by helium emission, the helium produced per DPA is much larger than in reactors. In fission reactors, about 0.5 appm is produced. This is negligible compared with the European Spallation Source, which is planned to operate with 2.5 GeV protons at a power of 5 MW. In this case there will be approximately 100 appm of helium per DPA produced in the steel window of the target [40]. In Ref. [40], a beam energy of 1.5 GeV was assumed. Since the inelastic cross-section is almost constant, the proton flux needed to achieve 5 MW will be reduced by roughly a factor of two (corresponding to the ratio of the beam energies). This means that only 50 instead of 100 appm/DPA of helium will be produced, which is still a very large number, as we will see below. To calculate the DPA per year, one needs the displacement cross-section, which is approximately 3000 barn. The proton flux integrated over one year is roughly $10^{22}$ protons/cm$^2$. Multiplying the number of protons by the displacement cross-section according to Eq. (9) leads to 30 DPA per year. After three years, about 2% of the material will be helium. This calculation is not realistic, however, since the steel window will break before it reaches a helium concentration of 2%. A limit of 10 DPA was estimated for the lifetime of the MEGAPIE target, which consisted of a steel vessel (T91 martensitic steel) filled with liquid lead–bismuth eutectic [41]. This was used in the 590 MeV continuous proton beam at the PSI, and therefore the conditions of the irradiation at the ESS will be different. As has already been explained, such predictions can depend on many parameters.

It should be mentioned that about 10 times more hydrogen than helium is formed during the irradiation. Hydrogen is even more mobile than helium but it usually leaves the material, except in the case of metals which form hydrides. In this case hydrogen embrittlement is an issue, and occurs even at temperatures from $0.1T_m$ to $0.4T_m$. Above this temperature range, the bonding in the hydride becomes unstable and the hydrogen can leave the metal.

Owing to the dependence on several parameters, a large number of experiments have to be performed to make predictions for any specific case. In particular, not much data for high-energy accelerator particles is available. At present, it is not at all clear how to transfer the database obtained from irradiation with low-energy neutrons to high-energy particle beams. The solution of this very complex problem cannot come from theoretical considerations alone; irradiation test experiments are needed. The good news is that the displacement cross-section becomes constant at higher energies, and therefore experiments at energies of about 1 GeV are sufficient. The results can then be used over a wider range of energies.

## 9 Summary

In this paper, two subjects were presented, the activation of and damage to materials, which are both caused by reactions between incident particles and target nuclei. The predictive power of Monte Carlo simulations for activation is usually limited by the knowledge of the production cross-sections. Most of the reaction cross-sections have to be provided by theories, owing to the lack of experimental data. However, more information is available for common dose-relevant isotopes. Therefore the prediction of the dose for these is usually much more accurate than that of the amount of activity for an exotic isotope. In practical applications, another large source of uncertainty is the material composition. The activity of isotopes which are produced by direct reactions, for example neutron capture, is proportional to the abundance of the element from which they are produced. A knowledge of the nuclide inventory is required for the disposal of radioactive waste and also for the transport of radioactive material. The radioactive waste from accelerators consists mainly of steel with a low to intermediate level of radioactivity. This is different from the situation for nuclear power plants, where the amount of steel is much lower than the total volume of waste.

Often, radiation damage leads to the failure of a component which is being used in a harsh environment. As accelerators become increasingly powerful, the understanding of radiation damage and the prediction of the lifetime of components becomes increasingly important. Unfortunately, the mechanisms leading to radiation damage are difficult to predict from theory alone. In addition, their effect depends on many parameters, such as the temperature and the type and energy of the particles. Higher temperatures and higher energies lead to a higher rate of recombination of defects, i.e. healing. The DPA is widely used as a measure of irradiation doses leading to radiation damage. Unfortunately, its value can depend strongly on the code used for calculation. The relation between irradiation conditions and macroscopic effects such as hardening and loss of thermal conductivity is not yet understood. Since experimental data taken at accelerator energies are scarce, more irradiation stations are needed to perform experiments under different conditions, the result of which can then be compared.

**Acknowledgement**

I would like to thank Sabine Teichmann and Yong Dai for reading the manuscript.